\pdfoutput=1
\documentclass[aps,prb,reprint,noshowpacs,letterpaper,floatfix]{revtex4-1}

\usepackage{amsmath,amssymb}
\usepackage{mathtools}        % \prescript
\usepackage{bm}
\usepackage{textcomp}
\usepackage{lmodern}
\usepackage[pdftex]{graphicx}

\begin{document}

\title{Single ferromagnetic fluctuations in UCoGe revealed by $^{73}$Ge- and
$^{59}$Co-NMR studies}

\author{Masahiro Manago}
  \email{manago@scphys.kyoto-u.ac.jp}
\author{Kenji Ishida}
  \affiliation{Department of Physics, Graduate School of Science,
  Kyoto University, Kyoto 606-8502, Japan}
\author{Dai Aoki}
  \affiliation{IMR, Tohoku University, Oarai, Ibaraki 311-1313, Japan}
  \affiliation{INAC/PHELIQS, CEA-Grenoble, 38054 Grenoble, France}

\begin{abstract}
$^{73}$Ge and $^{59}$Co nuclear magnetic resonance (NMR) and nuclear quadrupole
resonance (NQR) measurements have been performed on a $^{73}$Ge-enriched
single-crystalline sample of the ferromagnetic superconductor UCoGe in the
paramagnetic state.
The $^{73}$Ge NQR parameters deduced from NQR and NMR are close to
those of another isostructural ferromagnetic superconductor URhGe.
The Knight shifts of the Ge and Co sites are well scaled to each other
when the magnetic field is parallel to the $b$ or $c$ axis.
The hyperfine coupling constants of Ge are estimated to be close to
those of Co.
The large difference of spin susceptibilities between the $a$ and $b$
axes could lead to the different response of the superconductivity
and ferromagnetism with the field parallel to these directions.
The temperature dependence of the nuclear spin-lattice relaxation rates
$1/T_1$ at the two sites is similar to each other above 5 K.
These results indicate that the itinerant U-$5f$ electrons are responsible for
the ferromagnetism in this compound, consistent with previous studies.
The similarities and differences in the three ferromagnetic superconductors are
discussed.
\end{abstract}

\maketitle

\section{Introduction}

Uranium-based ferromagnetic (FM) super\-conductors\cite{JPSJ.81.011003,%
JPSJ.83.061011,JPSJ.76.051011,JPSJ.83.061012,PhysicaC.514.368}
have attracted much attention because of the intimate
relationship between ferromagnetism and superconductivity.
In these systems, the superconducting (SC) phases are inside the FM state,
and both of the ordered states are attributed to itinerant U-$5f$ electrons.
The pairing state and SC mechanism have been considered to be different from
the ordinary $s$-wave pairing mediated by the electron-phonon coupling.

UCoGe is a member of the FM superconductors, and has a Curie
temperature $T_\text{Curie} \simeq 3$ K and SC transition temperature
$T_\text{SC} \simeq 0.7$ K.\cite{PhysRevLett.99.067006}
We have shown the relationship between FM fluctuations with the Ising
anisotropy and superconductivity from the measurements of the field-angle and
field-magnitude dependencies, and suggested that the FM fluctuations induce
spin-triplet superconductivity in UCoGe.\cite{PhysRevLett.108.066403,%
JPSJ.83.073708}
Recently, Wu \textit{et al.} showed that this scenario explains the
macroscopic properties of superconductivity in UCoGe
quantitatively.\cite{NatComm.8.14480}
An interesting and important question in the FM superconductors is whether
this scenario is applicable for UGe$_2$ and URhGe.
To answer the question, we need to know similarities and differences of the
magnetic properties of three FM superconductors.

The superconductivity of UCoGe is enhanced
at the FM critical pressure $P_\text{c} \simeq 1$ GPa,
and the SC phase persists even if the ferromagnetism is
suppressed by the pressure\cite{JPSJ.77.073703,PhysRevLett.103.097003},
and this phase disappears at $P \simeq 4$ GPa.\cite{PhysRevB.94.125110}
In contrast, the SC state in UGe$_2$ terminates at the FM critical
pressure.\cite{Nature.406.587}
In addition, it seems that the relationship between magnetic properties and
superconductivity is also different between URhGe and UCoGe with respect to
the enhancement of superconductivity by the field parallel to the $b$ axis
in the orthorhombic structure.\cite{Science.309.1343,JPSJ.78.113709}
A reorientation of the magnetic moment occurs in URhGe at
$\mu_0 H \sim 12$ T,\cite{Science.309.1343}
which is most likely related to the strong
FM fluctuations parallel to the $b$ axis,\cite{PhysRevLett.114.216401} and
the re-entrant SC phase is observed in the limited field region around 12 T.
In contrast, the above picture that the FM moment rotates under the field does
not apply to the enhancement of superconductivity at $\sim 12$ T in UCoGe,
since such a moment polarization occurs at $\sim 50$ T along the
$b$ axis.\cite{PhysRevB.86.184416}
Magnetic characters of three FM superconductors have some differences against
the pressure and field responses, although they also have intimate
similarities, for instance, that their ferromagnetism is in the itinerant
regime from the angle resolved photoelectron spectroscopy (ARPES).%
\cite{PhysRevB.89.104518,PhysRevB.91.174503}
To understand such similarities and differences more precisely,
it is crucial to investigate magnetic characters from the same probes
throughout these systems.
The $^{73}$Ge-nuclear magnetic resonance (NMR) and nuclear quadrupole
resonance (NQR) measurements are valuable since Ge are included in all
the three compounds, and make it possible to compare these FM superconductors
from the microscopic point of view.
The $^{73}$Ge NMR and NQR have been performed on UGe$_2$%
\cite{JPSJ.74.705,JPSJ.74.2675,PhysRevB.75.140502}
and URhGe\cite{JPSJ.84.054710} in the
previous studies, and in this paper, we report first $^{73}$Ge NMR and NQR
results on UCoGe.

We also performed $^{59}$Co NMR in the same sample of UCoGe for
clarifying how the U-$5f$ electrons of UCoGe interact with the $^{59}$Co and
$^{73}$Ge nuclei.
Previously, we have shown from $^{59}$Co NMR and band calculation on UCoGe
as well as a reference compound YCoGe that the Co-$3d$ state is not in the
Fermi level, and is in nonmagnetic state.\cite{JPSJ.80.064711}
However, there are several reports that the hybridization between U-$5f$ and
Co-$3d$ is strong, and the Co-$3d$ contributes the magnetic moments.%
\cite{PhysRevB.91.174503,PhysRevB.92.121107}
We consider that the comparison between the hyperfine field at the Co and Ge
sites gives valuable information about the hybridization between
U-$5f$ and Co-$3d$.

In this paper, we show from the comparison between the $^{59}$Co and $^{73}$Ge
NMR and NQR that the magnetic fluctuations are governed by the
single-component fluctuations from the U-$5f$ electrons,
implying that those arising from the Co-$3d$ electrons are negligibly small
below 3 T.
These results are in good agreement with previous studies.\cite{JPSJ.80.064711}
We also found that the spin susceptibility
along the $a$ axis is substantially smaller than that along the $b$ axis.
This large difference may have some relation
to the different field responses of superconductivity and magnetism
along these directions.\cite{JPSJ.78.113709,PhysRevB.86.184416}
In addition, we show from the comparison of $^{73}$Ge-NQR results in the three
FM superconductors that their magnetic fluctuations are similar, but the
itinerant degree of U-$5f$ electron is different, which is consistent with the
transport behavior and magnitude of the ordered moment.%
\cite{JPSJ.81.011003,JPSJ.83.061011}

\section{Experiment}

\begin{figure}
	\centering
	\includegraphics{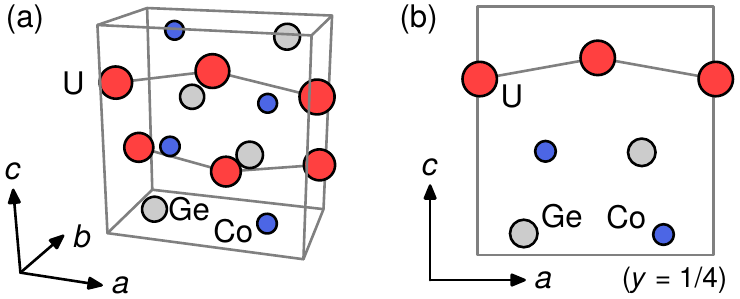}
	\caption{\label{fig:unitcell}(Color online)
	(a) A crystal structure of UCoGe.
    The largest, middle, and the smallest circles indicate U, Ge, Co atoms,
    respectively.
    (b) Atomic positions at $y = 1/4$ plane.
	}
\end{figure}

Crystal structure of UCoGe is shown in Fig.~\ref{fig:unitcell}.
UCoGe has a crystal structure of $\mathit{Pnma}$ space group
(\#62, $D_{2h}^{16}$),\cite{JAlloysCompd.234.225} and the local symmetry
of Co and Ge (and also U) atoms are expressed by $[.m.]$.

\begin{table}[htp]
    \caption{\label{table:CoGe_parameter}The data of $^{59}$Co and $^{73}$Ge
    nuclei; the nuclear spin $I$, the nuclear gyromagnetic ratio
    $\gamma_\text{n}$,
    the nuclear quadrupolar moment $Q$, and the natural abundance (N.A.).
    }

\begin{tabular}{rcccc}\hline\hline
& $I$ & $\gamma_\text{n}/2\pi$ (MHz/T)\footnote{Ref.~\onlinecite{Carter}}
& $Q$ ($10^{-28}$ m$^2$)\footnote{Ref.~\onlinecite{AtData.111.1}}
& N.A. (\%)\footnote{Ref.~\onlinecite{PAC.75.683}}\\ \hline
$^{59}$Co & $7/2$ & 10.03 & $+0.42(3)$ & 100 \\
$^{73}$Ge & $9/2$ & 1.4852 & $-0.196(1)$ & 7.76(8) \\ \hline\hline
\end{tabular}
\end{table}

We used a $^{73}$Ge-enriched single-crystalline UCoGe sample in this study.
The resistivity shows a broad hump at around 1.3 K, suggestive of a FM
transition.
From resistivity and susceptibility measurements, this sample exhibits a SC
transition at $T_\text{SC} \simeq 0.48$ K.
These transition temperatures are lower than those of the previous
higher quality samples [for instance,
$T_\text{SC} \simeq 0.57$ K (Ref.~\onlinecite{JPSJ.79.023707}) and
$T_\text{Curie} \simeq 3$ K (Ref.~\onlinecite{PhysRevLett.99.067006})].
The lower $T_\text{SC}$ and $T_\text{Curie}$ are due to the quality of the
enriched Ge ingredient.

We have performed NMR and NQR measurements at the Ge and Co sites down to
1.5 K to investigate spin susceptibility and FM fluctuations, and their
anisotropy in the paramagnetic state.
The nuclear parameters of $^{59}$Co and $^{73}$Ge nuclei, for which NMR and
NQR are possible, are listed in Table~\ref{table:CoGe_parameter}.
The NMR measurement was performed with a split-coil SC magnet
with a single-axis rotator.
The field was applied parallel to the crystallographic $a$, $b$, and $c$
axes, and the directions were determined from the $^{59}$Co NMR spectra.
The details of the alignment of the single-crystalline sample were described
in a previous paper.\cite{JPSJS.80SA.SA007}
NMR and NQR spectra depend on the electric field gradient (EFG) at the
nuclear sites, and the EFG tensor has three principal axes.
Usually, the EFG are represented with the NQR frequency
$\nu_\text{Q} \propto V_{zz}$ and asymmetric parameter $\eta \equiv
\lvert (V_{xx} - V_{yy})/ V_{zz}\rvert$, where $V_{ii}$ ($i=x,y,$ and $z$)
is an eigenvalue of the EFG tensor along the principal axis $i$ and
$\lvert V_{zz} \rvert \ge \lvert V_{yy} \rvert \ge \lvert V_{xx} \rvert$.
The direction $z$ is referred as ``maximum principal axis'' hereafter.
In the case of Co and Ge in UCoGe, one of the principal axes is parallel
to the $b$ axis from the local symmetry, but a degree of freedom remains
in the other principal-axis directions.

\section{Results and Discussion}

\begin{figure}
	\centering
 	\includegraphics{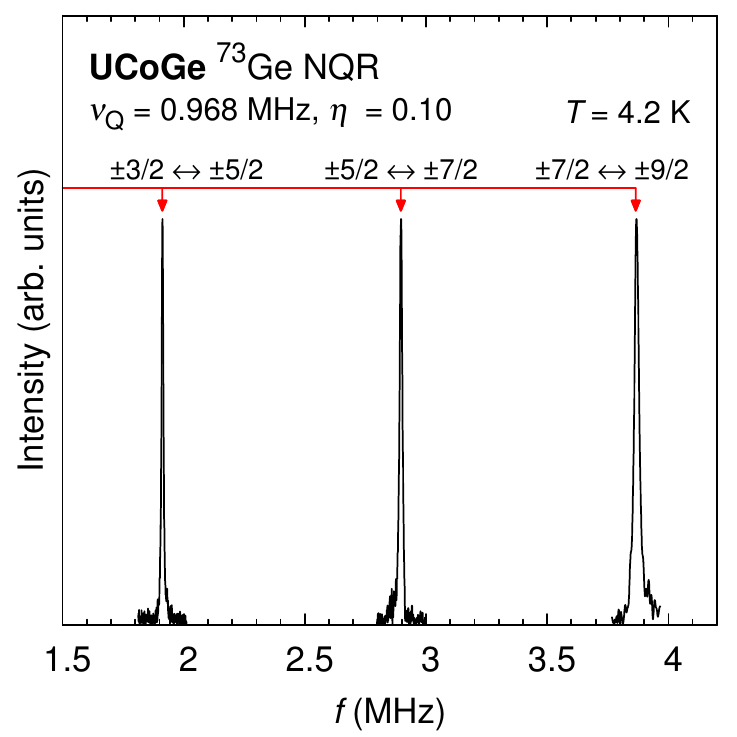}
	\caption{\label{fig:spectrum}(Color online)
	$^{73}$Ge NQR spectrum in UCoGe under zero magnetic field
	at 4.2 K (paramagnetic state).
	Three out of four peaks expected in $I=9/2$ are observed.
	The NQR parameters are deduced as $\nu_\text{Q} = (0.968 \pm 0.001)$ MHz
	and $\eta = 0.10 \pm 0.01$.
	The arrows represent the calculated peak positions
	with these parameters.
	}
\end{figure}

\begin{table}[htp]
\caption{\label{table:nqrparameter}The $^{73}$Ge and $^{59}$Co NQR parameters
in UCoGe and URhGe;
the NQR frequency $\nu_\text{Q}$, asymmetric parameter $\eta$, and the angle
$\theta_{zz}$ between the crystal $a$ axis and the maximum principal axis.
The maximum principal axis lies in the $ac$ plane at all the sites.}
\begin{tabular}{rccccc}\hline\hline
& & $\nu_\text{Q}$ (MHz) & $\eta$ & $\theta_{zz}$ (\textdegree) & Ref.
\\ \hline
UCoGe & $^{73}$Ge & $0.968 \pm 0.001$ & $0.10 \pm 0.01$ & 15.5 & this work \\
      & $^{59}$Co & 2.85 & 0.52 & 10   & \onlinecite{JPSJ.79.023707} \\
URhGe & $^{73}$Ge & 1.06 & 0.09 & 9.8 & \onlinecite{JPSJ.84.054710} \\
\hline\hline
\end{tabular}
\end{table}

Figure \ref{fig:spectrum} shows the $^{73}$Ge-NQR spectrum in UCoGe at 4.2 K
at zero field.
$^{73}$Ge has $I=9/2$ nuclear spin, and the observed three peaks
correspond to the $\pm 3/2 \leftrightarrow \pm 5/2$,
$\pm 5/2 \leftrightarrow \pm 7/2$, and $\pm 7/2 \leftrightarrow \pm 9/2$
transitions.
The lowest peak corresponding to $\pm 1/2 \leftrightarrow \pm 3/2$
transitions could not be detected due to the low frequency
beyond the range of our NMR receiver.
The $^{73}$Ge quadrupole frequency $\nu_\text{Q}$ and the asymmetric parameter
$\eta$ of UCoGe at $T=4.2$ K are determined from the experiments as shown in
Table~\ref{table:nqrparameter}.
These parameters are close to those of URhGe,\cite{JPSJ.84.054710}
but $\eta$ is much smaller than that at the Co site in UCoGe.%
\cite{JPSJ.79.023707}

\begin{figure}
	\centering
	\includegraphics{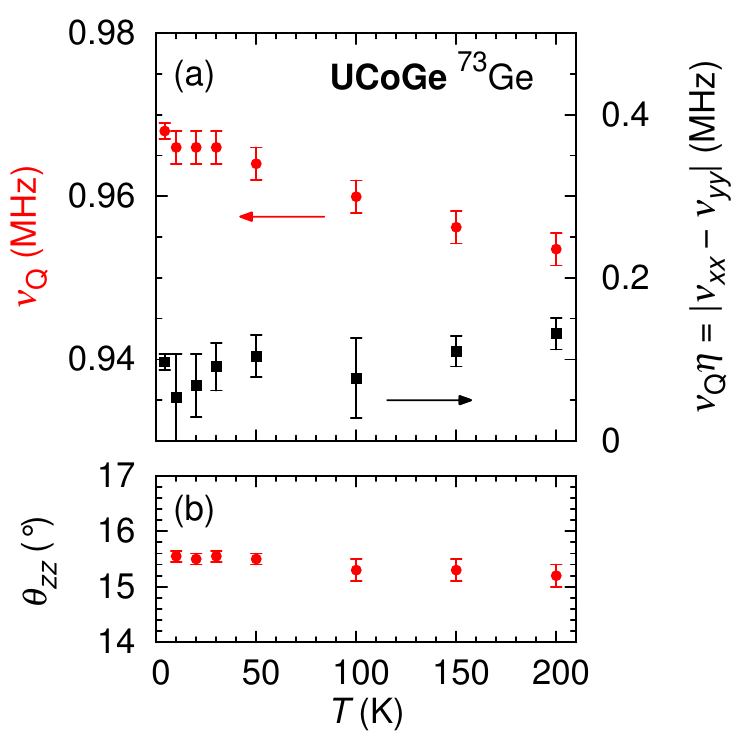}
	\caption{\label{fig:nuq}(Color online)
	Temperature dependence of EFG parameters of $^{73}$Ge.
    (a) $\nu_\text{Q}$ and $\nu_\text{Q} \eta = \lvert \nu_{xx} - \nu_{yy}
    \rvert$,
    (b) the angle $\theta_{zz}$ between the maximum principal axis of EFG and
    the crystallographic $a$ axis in the $ac$ plane are shown.
    $\nu_\text{Q}$ and $\eta$ are determined from the NQR spectra at $H=0$,
    while $\theta_{zz}$ is deduced from the field-swept NMR spectra.
	}
\end{figure}

The temperature dependence of $\nu_\text{Q}$ and $\nu_\text{Q} \eta$
are shown in Fig.~\ref{fig:nuq} (a),
and $\nu_\text{Q}$ slightly decreases as temperature increases.
At finite temperature, NQR frequency in metals is empirically expressed
as\cite{ZPhysB.24.177}
$\nu_\text{Q} (T) = \nu_\text{Q} (0) (1 - \alpha T^{3/2})$,
where $\alpha$ is a positive value.
Although the present result is not so accurate to distinguish
whether this relation holds, the monotonic decrease of $\nu_\text{Q}$
with increasing temperature is a conventional behavior.
No large temperature variation of the anisotropic parameter
was detected in this system.

\begin{figure}
	\centering
	\includegraphics{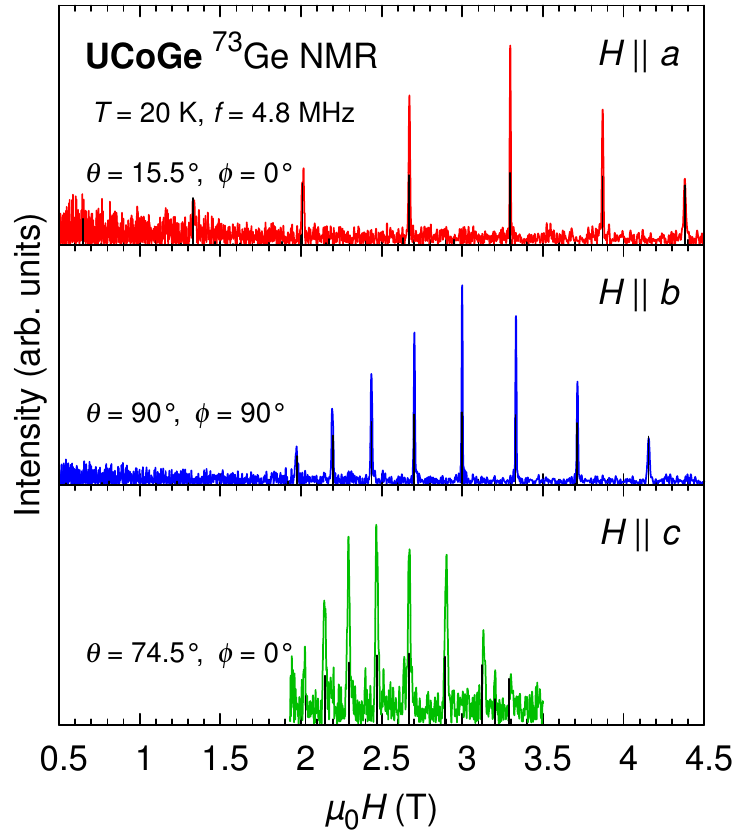}
	\caption{\label{fig:hsp}(Color online)
	Field-swept $^{73}$Ge NMR spectra in UCoGe with a fixed frequency
	$f = 4.8$ MHz at 20 K for three field directions.
	The vertical lines indicate the calculated peak positions (see text).
	The polar angle $\theta$ and the azimuthal angle $\phi$ represent
	the field directions with respect to the coordinate of the electric field
	gradient [$(\theta, \phi) = (90\text{\textdegree}, 0\text{\textdegree})$
    corresponds to $V_{xx}$ direction].
	}
\end{figure}

Figure \ref{fig:hsp} shows the field-swept spectra of $^{73}$Ge along
the three directions.
The principal axes of EFG at the Ge site are deduced from these spectra.
The best fit to the experimental data was obtained when
the maximum principal axis is in the $ac$ plane and tilts
$\theta_{zz} = 15.5$\textdegree\ from the $a$ axis at 20 K,
and the second principal axis is parallel to the $b$ axis.
The temperature dependence of $\theta_{zz}$ is shown in
Fig.~\ref{fig:nuq} (b).
This change seems to be tiny, but it cannot be neglected for extracting the
accurate Knight shift of $^{73}$Ge owing to the small gyromagnetic ratio.
Thus, the Knight shift is calculated after subtracting the
temperature-dependent EFG for the $^{73}$Ge site.
It is interesting that the maximum principal axes of EFG at the
Co\cite{PhysRevLett.105.206403} and Ge sites
in UCoGe and the Ge site in URhGe\cite{JPSJ.84.054710}
are roughly parallel to the $a$ axis.
This feature may originate from the crystal structure because that of UCoGe
and URhGe can be regarded as a deformed hexagonal AlB$_2$-type structure,
and the hexagonal $c$ axis, which is the maximum principal axis,
corresponds to the orthorhombic $a$ axis in UCoGe and URhGe.

\begin{figure}
	\centering
	\includegraphics{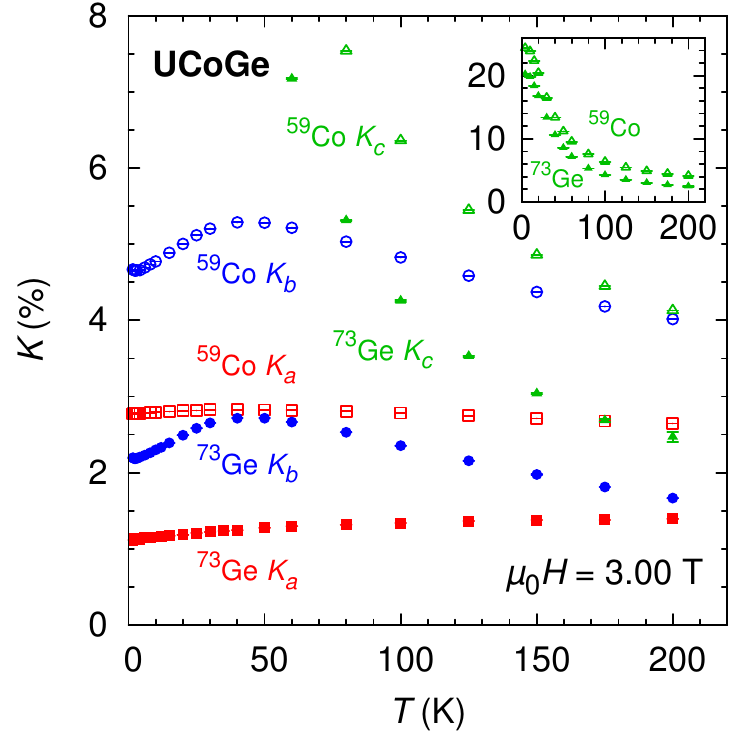}
	\caption{\label{fig:knightshift}(Color online)
	$^{73}$Ge (closed symbol) and $^{59}$Co (open symbol)
	Knight shifts measured at the central line ($1/2 \leftrightarrow -1/2$)
	with the field of 3 T
	parallel to the $a$ (squares), $b$ (circles), and $c$ (triangles) axes.
	The inset shows the result of $c$ direction with a different scale.
	}
\end{figure}

Figure \ref{fig:knightshift} shows the $^{73}$Ge and $^{59}$Co Knight shifts
of three directions with a fixed field of 3 T at the central line
($1/2 \leftrightarrow -1/2$).
The Knight shift in UCoGe is highly anisotropic with the $c$ direction being
an easy axis, resulting from the strong Ising anisotropy.

The Knight shift along the $i$ direction ($i$ = $a$, $b$, and $c$) at the $m$
site ($m$ = 73 and 59) is described as
\begin{equation}\label{eq:knightshift}
\prescript{m}{}{K}_{i} = \prescript{m}{}{A}_{i} \chi_{\text{spin},i}
+ \prescript{m}{}{K}_{\text{orb},i},
\end{equation}
where $\prescript{m}{}{A}_{i}$ is the hyperfine coupling constant,
$\chi_{\text{spin},i}$ is the spin susceptibility,
and $\prescript{m}{}{K}_{\text{orb},i}$ is the orbital part of
the Knight shift.
The latter part is usually temperature independent in $d$-electron systems,
because a crystal electric field (CEF) splitting is much larger than room
temperature,
while it is temperature dependent in $f$-electron systems, where the
CEF is not so large as in the $d$-electron systems.
We also note that the first term of the right hand side of
Eq.~(\ref{eq:knightshift}) is no longer pure spin in the $f$-electron
system because of the strong spin-orbit interaction,
and this indicates the susceptibility of quasiparticles.\cite{JPSJ.74.1245}
Nevertheless, we use the term ``spin susceptibility'' for simplicity.
\begin{figure}
	\centering
	\includegraphics{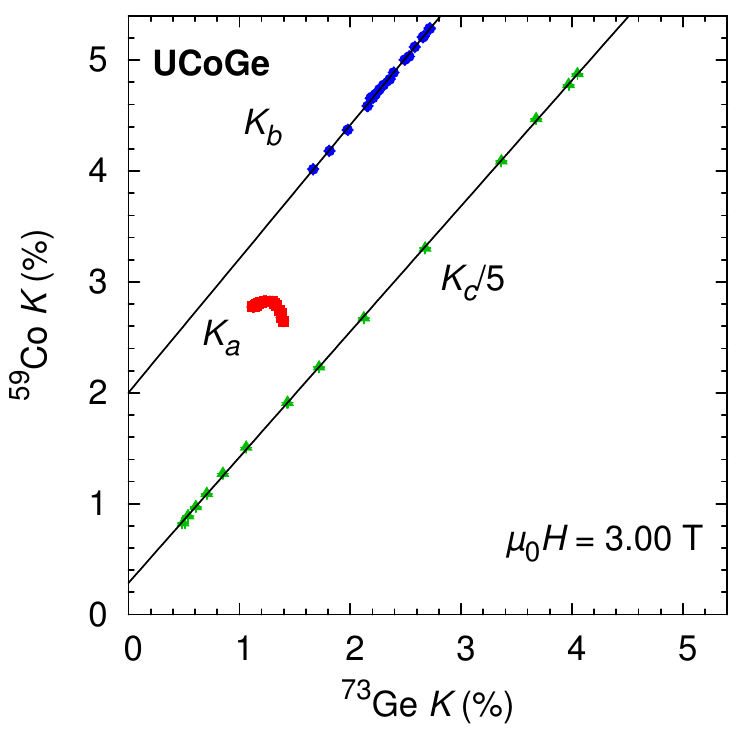}
	\caption{\label{fig:k-k}(Color online)
	$^{59}$Co Knight shifts versus those of $^{73}$Ge of three
	directions with the temperature being an implicit parameter.
	The Knight shifts along the $c$ axis are scaled to $1/5$.
	The solid lines are the best fit of the linear relation of the Knight
	shifts for $H \parallel b$ and $c$.
	}
\end{figure}

Figure \ref{fig:k-k} shows the relation of Knight shifts between the Ge and Co
sites, where temperature is an implicit parameter.
When the field is parallel to the $b$ or $c$ axis,
good linear relations hold between two sites in a wide temperature range.
This result indicates that the dominant temperature dependence of the Knight
shift is attributed to the single component of the spin susceptibility from
the U-$5f$ electrons, which is in good agreement with the previous reports
that the magnetism is carried by U.%
\cite{PhysRevB.92.035124,PhysRevB.92.121107}
The good linearity also implies that the simple treatment of the Knight shifts
described as Eq.~(\ref{eq:knightshift}) is valid even in the $5f$ electron
systems since the system has a large spin susceptibility and the temperature
dependence of $K_\text{orb}$ is relatively small.
The hyperfine coupling constants of $^{73}$Ge are estimated from the slopes of
the lines showing in Fig.~\ref{fig:k-k} and the hyperfine coupling constants
of $^{59}$Co reported previously,\cite{PhysRevLett.105.206403}
and they are $\prescript{73}{}{A}_b \simeq (4.3 \pm 0.1)$
$\text{T}/\mu_\text{B}$ and
$\prescript{73}{}{A}_c \simeq (4.2 \pm 0.1)$ $\text{T}/\mu_\text{B}$
along the $b$ and $c$ directions, respectively.
These values are 0.8--0.9 times those at $^{59}$Co, suggesting that the
U-$5f$ electrons couple to the $^{59}$Co and $^{73}$Ge nuclei almost equally.
In addition, if we assume that $\prescript{73}{}{K}_{\text{orb},i}
\sim 0.2$\%, which is a
typical value of the orbital shift of Ge and similar $p$-electron atoms such
as Ga and As\footnote{For instance, $K_\text{orb}$ of $^{71}$Ga(1) site in the
heavy-fermion superconductor PuCoGa$_5$ is
$\sim 0.1$\%,\cite{Nature.434.622} and that of $^{75}$As in the metallic
antiferromagnet BaFe$_2$As$_2$ is around $\sim 0.2$\%.\cite{JPSJ.77.114709}}
and is an order of magnitude smaller than
$\prescript{59}{}{K}_{\text{orb},i}$, then
$\prescript{59}{}{K}_{\text{orb},b}$ and
$\prescript{59}{}{K}_{\text{orb},c}$ are estimated to be
2.2 and 1.7\%, respectively.
These values are similar to $\prescript{59}{}{K}_\text{orb} \simeq 1.6$\%
in a non-magnetic metal YCoGe.\cite{JPSJ.80.064711}

In contrast, when the field is parallel to the $a$ axis, the temperature
dependence of the Knight shifts is relatively small, and the linear
relation is not seen between two sites.
These results suggest that the spin susceptibility along the $a$ axis
is much smaller than that of $b$ and $c$ axis since $^{m}A_i$ is considered
to be isotropic in this system.\cite{PhysRevLett.105.206403}
A possible origin of the anomalous $K_a$ is the temperature-dependent
$K_\text{orb}$ owing to the small CEF splitting, as mentioned before.
The $^{59}$Co Knight shift of the $a$ direction has a broad maximum at
$T^* \sim 40$ K as $^{59}K$ of the $b$ direction, while the $^{73}$Ge Knight
shift monotonically decreases with decreasing temperature with a broad kink
around $T^*$.
These anomalies may be related to the broad maximum of the Knight shifts and
bulk susceptibility\cite{PhysRevB.86.184416} in $H \parallel b$.
As discussed later, the anomaly around $T^*$ is also recognized in the
nuclear spin-lattice relaxation rate $1/T_1$, which suggests that the system
becomes more itinerant below $T^*$.

\begin{figure}
	\centering
	\includegraphics{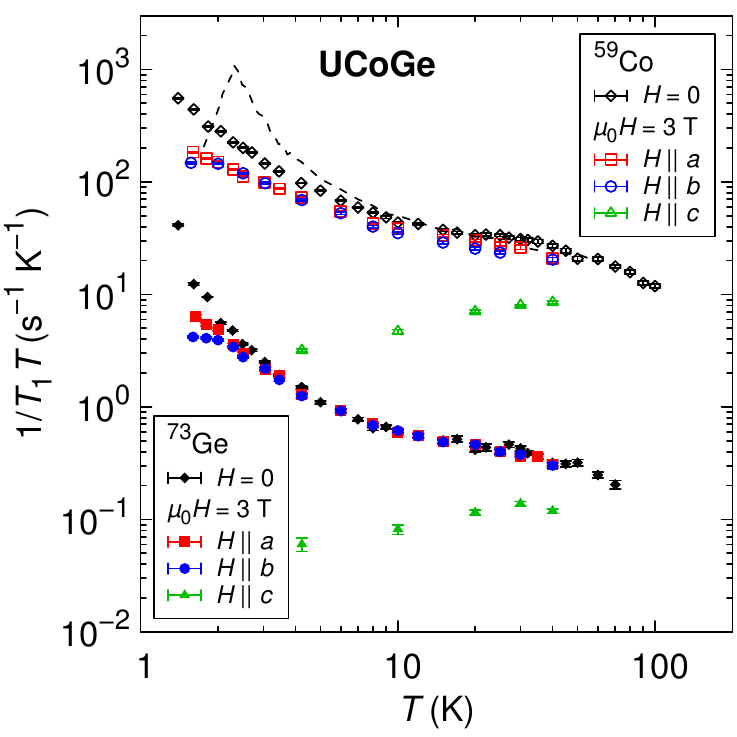}
	\caption{\label{fig:t1}(Color online)
	$^{73}$Ge (closed symbol) and $^{59}$Co (open symbol)
	nuclear spin-lattice relaxation rates
	$1/T_1$ divided by $T$ measured with NQR ($H=0$, diamond)
	and NMR ($\mu_0 H = 3$ T) with the field
	parallel to the $a$ (squares), $b$ (circles), and $c$ (triangles) axes.
	The dashed line indicates the previous $^{59}$Co NQR result with the
	different single-crystalline sample at the paramagnetic
	signal.\cite{JPSJ.79.023707}
	}
\end{figure}

Figure \ref{fig:t1} shows $1/T_1$
divided by temperature at $^{73}$Ge and $^{59}$Co sites under zero field
and fields of 3 T parallel to the $a$, $b$, and $c$ axes.
The dashed line shows $1/T_1T$ measured by the $^{59}$Co NQR in the previous
single-crystalline sample.\cite{JPSJ.79.023707}
Contrary to the previous result, the peak of $1/T_1$ showing the FM
transition was not detected down to 1.5 K in this sample.
When the field is parallel to the $a$ or $b$ axis, $1/T_1T$ is close
to that at zero field and is enhanced at low temperatures,
while it is strongly suppressed with $H \parallel c$.
This field-direction dependence of $1/T_1$ is consistent with the previous
results, which indicate that the FM fluctuations are strongly anisotropic%
\cite{PhysRevLett.105.206403} and are suppressed by
$H \parallel c$.\cite{PhysRevLett.108.066403}

\begin{figure}
	\centering
	\includegraphics{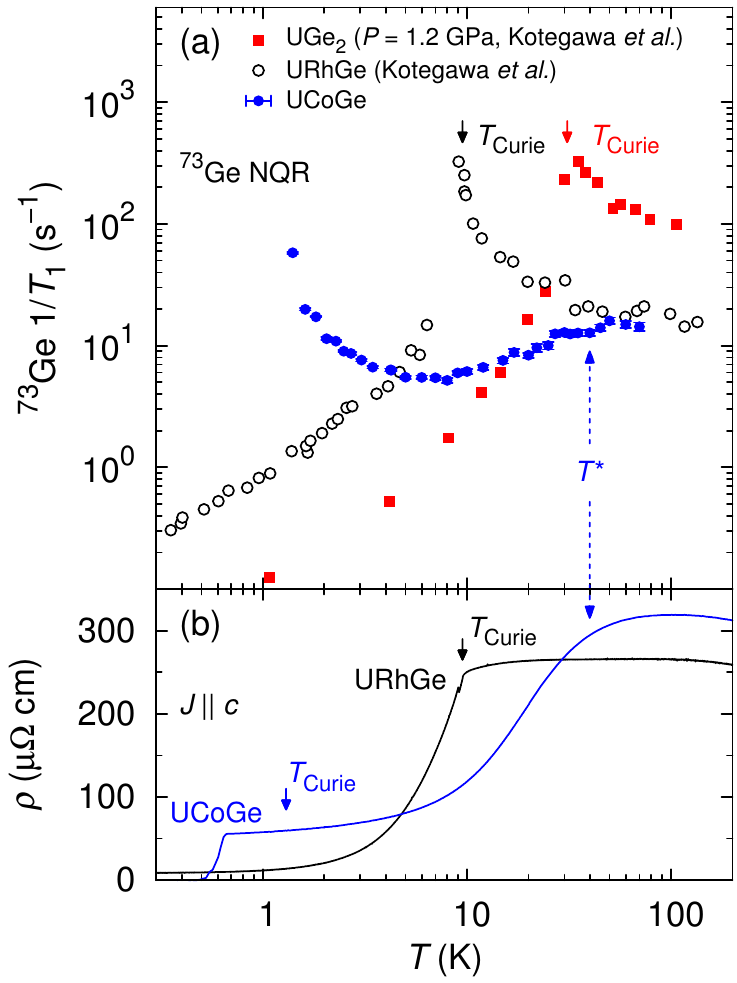}
	\caption{\label{fig:t1vs}(Color online)
	$1/T_1$ at $^{59}$Co divided by that at $^{73}$Ge measured by NQR and NMR.
	The dashed line is the expected relation (see text).}
\end{figure}

As shown in Fig.~\ref{fig:t1vs}, the behavior of $1/T_1$ at two sites is
essentially similar to each other above 5 K in any direction of the field,
although the deviation was observed in the lower temperatures, where the
FM fluctuations develop.
As discussed in the previous paper,
$1/T_1T$ at the $m$ site measured with $H \parallel \alpha$
is expressed in terms of the imaginary part of the dynamic susceptibility
along the $\beta$ and $\gamma$ directions
$\chi''_{\beta, \gamma}(\bm{q},\omega_0)$, perpendicular to $\alpha$,
as\cite{PhysRevLett.105.206403}
\begin{align*}
\prescript{m}{}{\left(\frac{1}{T_1T}\right)}_{\alpha} =
\frac{\prescript{m}{}{\gamma}_\text{n}^2 k_\text{B}}{(\gamma_\text{e}\hbar)^2}
\sum_{\bm{q}} \left[
\lvert \prescript{m}{}{A}_{\beta} \rvert^2
\frac{\chi''_{\beta}(\bm{q},\omega_0)}{\omega_0} \right. \\
\left. + \lvert \prescript{m}{}{A}_{\gamma} \rvert^2
\frac{\chi''_{\gamma}(\bm{q},\omega_0)}{\omega_0}\right],
\end{align*}
where $\prescript{m}{}{\gamma}_\text{n}$ and $\omega_0$ are the gyromagnetic
ratio at the $m$ site and NMR frequency, respectively.
The ratio of $1/T_1$ between $^{59}$Co and $^{73}$Ge site is expected to be
$(\prescript{59}{}{\gamma}_\text{n} / \prescript{73}{}{\gamma}_\text{n})^2
\cdot (\prescript{59}{}{A}/ \prescript{73}{}{A})^2 \simeq 59$,
shown by the dashed line in Fig.~\ref{fig:t1vs}, if the magnetic fluctuations
consist of a single component.
It is noted that the ratio estimated from the NMR measurements is close to
the expected one in any field direction,
indicating that single magnetic fluctuations arising from the U-$5f$ electrons
are dominant at both sites.
However, the ratio from the NQR measurements is slightly larger than the
expected value.
This is considered to be due the difference of the EFG parameters $\eta$ and
$\theta_{zz}$ as shown in Table \ref{table:nqrparameter}.
The relaxation curve in NQR is affected by $\eta$,\cite{JPhys.3.8103}
and the angle difference of the maximum principal axis
gives rise to the different weight of the FM fluctuations in $1/T_1$.

Below 5 K, the ratio becomes smaller with decreasing temperature, which is due
to the suppression of the increase of $1/T_1T$ at $^{59}$Co compared with that
at $^{73}$Ge.
As for the $1/T_1$ measurements with NQR, $H \parallel a$, and $H \parallel b$,
rf-pulse fields ($H_1$) were applied along the
$c$ axis, which is parallel to the direction of the Ising FM fluctuations.
In this case, we found that the value of $1/T_1$ near $T_\text{Curie}$
depends on the intensity of $H_1$, the smaller $H_1$ gives the larger
$1/T_1$.
Therefore, the deviation from the expected ratio would be due to the
difference of the effect of $H_1$ for the NMR measurements
between two nuclear sites.
When we compare $1/T_1$ of $^{73}$Ge with that of $^{59}$Co, we should be
careful for the presence of such differences.

\begin{figure}
	\centering
	\includegraphics{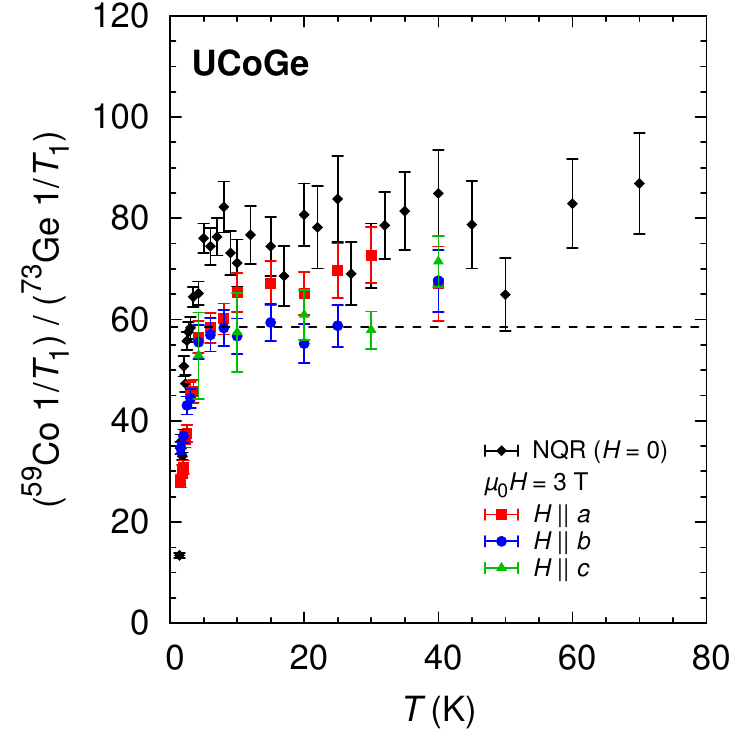}
	\caption{\label{fig:fst1}(Color online)
	(a) Temperature dependence of $1/T_1$ of $^{73}$Ge
	in the present sample of UCoGe (closed circle),
	UGe$_2$ at $P$ = 1.2 GPa (closed square),\cite{JPSJ.74.705}
	and URhGe (open circle)\cite{JPSJ.84.054710} under zero field.
	The arrows indicate $T_\text{Curie}$ for UGe$_2$ and URhGe, and $T^*$
	for UCoGe.
    (b) Temperature dependence of the electric resistivity in $J \parallel c$
    in the present sample of UCoGe and a single-crystal of URhGe.%
    \cite{D.Aoki}
	}
\end{figure}
Figure \ref{fig:fst1} shows the temperature dependence of $1/T_1$ of $^{73}$Ge
NQR in the present sample of UCoGe, along with that of UGe$_2$
($P = 1.2$ GPa)\cite{JPSJ.74.705} and URhGe\cite{JPSJ.84.054710} at $H=0$.
$1/T_1$ of $^{73}$Ge shows similar temperature dependence in three FM
superconductors, and, particularly, $1/T_1$ of UCoGe and URhGe becomes constant
with the similar values at higher temperatures.
This implies that the U-$5f$ is a localized state at higher
temperatures since in general $1/T_1$ is temperature independent in compounds
with local moments.\cite{PTP.16.23}
However, we note that the development of the FM fluctuations and FM ordering
in UCoGe occur after the gradual decrease of $1/T_1$ below $T^*$, where the
magnetic susceptibility $\chi$ deviates from the Curie-Weiss
behavior\cite{JPSJ.83.061012} and the electrical resistivity along the
$c$-axis shows metallic behavior as shown in Fig.~\ref{fig:fst1} (b).
As pointed out in the previous study, these behaviors are quite different from
those of URhGe.\cite{JPSJ.84.054710}
In URhGe, the development of the FM fluctuations and FM ordering occurs
where the most U-$5f$ is still in the localized state, which is
known from the $1/T_1$ and the electrical-resistivity behaviors shown in
Fig.~\ref{fig:fst1}.
Thus, these results clearly indicate that the itinerant degree of the U-$5f$
is different in UCoGe and URhGe, as discussed in
Ref.~\onlinecite{JPSJ.83.061012}, although the quasiparticle bands with large
contributions from U-$5f$ state were observed from
ARPES.\cite{PhysRevB.89.104518,PhysRevB.91.174503}
We suggest that the difference of the itinerant degree is one of the key
factors to understand the differences of the superconductivity and
ferromagnetism of these compounds.
Another interesting difference between UCoGe and URhGe is the anisotropy of
the spin susceptibility at low temperatures.
As shown in Fig.~\ref{fig:knightshift}, the strong Ising-type anisotropy with
the $c$ axis being the easy axis was observed in UCoGe,
but the susceptibilities
along the $b$ and $c$ axes are comparable above $T_\text{Curie}$ in URhGe.%
\cite{JPSJ.84.054710}
These differences are important to understand the differences of the
metamagnetic behavior and field-enhanced superconductivity observed in
URhGe\cite{Science.309.1343}
and UCoGe.\cite{PhysRevB.86.184416,JPSJ.78.113709}

The anisotropy of the spin susceptibility perpendicular to the Ising axis is
considered to be an origin of the different field responses between the $a$ and
$b$ axes in the superconductivity and magnetism of UCoGe.%
\cite{JPSJ.78.113709,PhysRevB.86.184416}
The FM phase is suppressed by the field parallel to the  $b$ axis as well as
the SC phase are reinforced at around $\mu_0 H \sim 12$ T,
while the $T_\text{Curie}$ and the $A$ coefficient of the resistivity
are hardly changed by the field parallel to the $a$ axis.
Because the spin susceptibility along the $b$ axis is much larger than that
along the $a$ axis, it is expected that the field along the former axis affects
the ferromagnetism more seriously.
Such anisotropic field responses are also reported in the re-entrant SC region
of URhGe: the reorientation of the FM moment occurs and re-entrant SC phase
arises along the $b$ axis, and these anomalies are insensitive to the
field along the $a$ axis.\cite{Science.309.1343,JPhys.21.164211}
Thus, UCoGe and URhGe have some similarity concerning the field dependencies
of the SC and FM phases.
In addition, the anisotropy of the spin susceptibility should be taken into
account when determining the $d$ vector in the spin-triplet SC state
of these FM superconductors.
Spin components of the Cooper pairs are active and perpendicular to the
$d$ vector in spin-triplet superconductors.\cite{RevModPhys.47.331}
Thus, it is a future task to reveal how the anisotropy of the spin
susceptibility in the normal state affects the structure of the order
parameter of the SC state in these systems.

Finally, we comment on the relation between the FM and SC phases in UCoGe.
The present sample exhibits the FM transition at around $T_\text{Curie}
\sim 1.3$ K, which is much lower than $\sim 3$ K reported
previously.\cite{PhysRevLett.99.067006}
Since $T_\text{SC} = 0.48$ K is comparable to the previous results
[$T_\text{SC} = 0.57$ K (Ref.~\onlinecite{JPSJ.79.023707})],
it is considered that the FM phase is more sensitive to the quality of the
sample than the SC phase, as pointed out in a previous
study.\cite{JPSJ.80.084709}
It was reported that the SC phase can exist even without the static FM
ordering.\cite{PhysRevB.82.180517}
Since the SC phase without the FM phase is also induced by the hydrostatic
pressure,\cite{JPSJ.77.073703,PhysRevLett.103.097003} the FM ordering is not a
necessary condition of the SC phase in UCoGe but the FM fluctuations are,
as discussed in previously.\cite{PhysRevLett.108.066403,NatComm.8.14480}
This is also inferred from the pressure dependence of $T_\text{SC}$, which
exhibits no discontinuity across the FM transition line.%
\cite{JPSJ.77.073703,PhysRevLett.103.097003}
In this sense, UCoGe seems to be a typical example of the system
where the SC phase is induced by the FM quantum fluctuations.
Thus, the order of the FM transition appears to be the second order, although
the first-order-like behavior was observed in the previous NQR spectrum.%
\cite{JPSJ.79.023707}
Further studies are still needed to uncover the nature of the FM quantum
transition in UCoGe, and the detailed comparison of the NMR results obtained
in the different quality of samples will be summarized in a separated paper.

\section{Summary}

We performed the $^{73}$Ge and $^{59}$Co NMR and NQR measurements
on the paramagnetic state of UCoGe, and found that the electric field gradient
of the Ge site in UCoGe is close to that of the isostructural compound URhGe.
It was revealed that the static and dynamic spin susceptibilities at these
sites are essentially similar to each other, but the spin susceptibility
along the $a$ axis is extremely small.
This result indicates that the U-$5f$ electrons are the dominant origin of the
ferromagnetism in this system and couple to the $^{73}$Ge and $^{59}$Co nuclei
almost equally.
In addition, it was found that the contribution of Co $3d$ electrons probed
with $^{59}$Co NMR and NQR is negligibly small.
Therefore, we can safely say that the $^{73}$Ge NMR and NQR give the
essentially the same information about $5f$ electrons as those of $^{59}$Co
in UCoGe.

The authors would like to thank S. Kitagawa, Y. Tokunaga, T. Hattori,
H. Kotegawa, Y. Maeno, S. Yonezawa, N. K. Sato, J.-P. Brison, D. Braithwaite,
A. Pourret, C. Berthier, A. de Visser,
and Y. Kitaoka for valuable discussions and contribution to experiments.
One of the authors (MM) is a Research Fellow of
Japan Society for the Promotion of Science (JSPS).
This work was supported by Kyoto University LTM Center, and by
Grant-in-Aid for Scientific Research (Grant No.~JP15H05745),
Grant-in-Aids for Scientific Research on Innovative Areas ``J-Physics''
(Grants No.~JP15H05882, No.~JP15H05884, and No.~JP15K21732), and
Grant-in-Aid for JSPS Research Fellow (Grant No.~JP17J05509) from JSPS.

\bibliography{bibliography}

\end{document}